\documentclass[aps,showpacs]{revtex4}
\usepackage{epsf,latexsym}

\newcommand{\mb}[1]{\mbox{\scriptsize #1}}
\newcommand{\rhom}{\rho_{\mb{MRE}}}

\begin{document}

\title{Finite-size effects on the QCD phase diagram}
\author{O. Kiriyama}
\affiliation{Institute f\"ur Theoretische Physik, J.W. Goethe-Universit\"at, D-60439
Frankfurt am Main, Germany}
\author{T. Kodama}
\author{T. Koide}
\affiliation{Instituto de F\'{\i}sica, Universidade Federal do Rio de Janeiro, C.P.
68528, 21945-970 Rio de Janeiro, RJ, Brazil}

\begin{abstract}
We discuss the finite-size effects on the chiral phase transition in QCD. We
employ the Nambu-Jona-Lasinio model and calculate the thermodynamic
potential in the mean-field approximation. Finite-size effects on the
thermodynamic potential are taken into account by employing the multiple
reflection expansion. The critical temperature is lowered and the order of
the phase transition is changed form first to second as the size of the
system of interest is reduced.
\end{abstract}
\pacs{11.30.Rd,12.38.Lg,12.39.-x}
\maketitle

\section{Introduction}

The phase structure of QCD at high temperature $T$ and/or quark chemical
potential $\mu$ has attracted a great deal of interest in cosmology, compact
stars and heavy-ion collisions. During the last decades, significant
advances have been made in our understandings of the phase structure of hot
and/or dense quark matter. At the present time, it is widely accepted that
the QCD vacuum undergoes a phase transition (into a chirally symmetric phase
or a color-superconducting phase) at sufficiently high $T$ and $\mu$.

The studies of the phase structure of QCD so far have been exclusively devoted to
homogeneous quark matter in bulk. However, the QGP phase created in
relativistic heavy-ion collisions has a finite size and it is not obvious
that the size is large enough to allow to apply the thermodynamic limit. If
the size of the system of interest is not large enough, we need to take
account of the deviations from thermodynamic calculations. For instance, the
fluctuations of order parameter, induced by the finite-size effects, would
lower the critical temperature and change the order of phase transition.

There are many works which concern with finite-size effects on hadron
physics; lattice QCD calculations \cite%
{Fuku1,Fuku2,Bail,Gopie,QCDSF,Koma,Orht,SS}, the finite-size scaling
analyses \cite{FSScailing1,FSScailing2,FSScailing3}, plasmon \cite{True},
baryon number fluctuations \cite{Singh}, and so on \cite%
{Tho,Davi,Mus,Rischke,Kim,Zak1,Zak2,Sto,Amore1,He,Kogut,Wong,Fraga}. 
In particular, the finite-size effects on the QCD phase diagram are studied in \cite{He} and 
\cite{Kogut}. In the former, however, the discussion is limited to
(2+1)-dimensional systems and in the latter, the phase transition discussed
is not related directly with the manifestation of chiral symmetry.

In this paper, we discuss the finite-size effects to the chiral phase
transition. For the purpose of illustrating our approach, we adopt the
two-flavor Nambu--Jona-Lasinio (NJL) model as an effective model of QCD. The
finite-size effects are taken into account by using the multiple reflection
expansion (MRE) \cite{BB,BJ,Mad3}. Then, we study the size dependence of the
thermodynamic potential in the mean-field approximation.

In the MRE approximation, the finite-size effects are included in terms of
the modified density of states. The MRE has been used to calculate the
thermodynamic quantities (such as the energy per baryon and the free energy)
of finite lumps of quark matter \cite{Mad4}. As far as overall structures
are concerned, the results are in good agreement with those of the
mode-filling calculations with the MIT bag wave functions. We emphasize that
our model calculation differs from the finite-size scaling analyses (in
which finite-size effects act as an external field; more specifically, mass)
and the studies in a finite box in that the present model is essentially
based on the MIT bag wave functions. However, it should be noted that the
relevance of the MRE approximation is still inconclusive when it is applied
to nonperturbative calculations \cite{KH,K1,YHT,K2}.

This paper is organized as follows. In section 2, we apply the MRE
approximation to the NJL model and calculate the $T$-$\mu$ phase diagram.
However, the result is unacceptable because the order of chiral phase
transition is changed from second to first. In section 3, we reconsider the
application of the MRE to the chiral phase transition. Summary and
concluding remarks are given in section 4.

\section{simple application of MRE}

We consider up and down quarks to be massless and start with the following $%
\mathrm{SU}(2)_L\times\mathrm{SU}(2)_R$ symmetric NJL Lagrangian, 
\begin{eqnarray}
\mathcal{L}=\bar{q}i\gamma^{\mu}\partial_{\mu}q +G\left[(\bar{q}q)^2+(\bar{q}%
i\gamma_5\vec{\tau}q)^2\right],
\end{eqnarray}
where $q$ denotes the quark field with two flavors $(N_f=2)$ and three
colors $(N_c=3)$ and $G$ is the coupling constant. The Pauli matrices $\vec{%
\tau}$ act in the flavor space. The ultraviolet cutoff $\Lambda_{%
\mbox{\scriptsize UV}} = 0.65$ GeV and the coupling constant $G=5.01~\mathrm{%
GeV}^{-2}$ are determined so as to reproduce the pion decay constant $%
f_{\pi}=93$ MeV and the chiral condensate $\langle \bar{q}q \rangle = (-250~%
\mathrm{MeV})^3$ in the chiral limit \cite{Kle,HK}.

In the mean-field approximation, the thermodynamic potential per unit volume 
$\omega$ is given by 
\begin{eqnarray}
\omega = \frac{m^2}{4G}-\nu\int^{\Lambda_{\mbox{\scriptsize UV}}}_0
dk\rho(k)\left\{ E_k +\frac{1}{\beta}\ln\left[1+e^{-\beta(E_k + \mu)}\right] %
\left[1 + e^{-\beta(E_k - \mu)}\right] \right\},  \label{eqn:TDP1}
\end{eqnarray}
where $\nu=2N_cN_f$, $\beta=1/T$ and $E_k = \sqrt{k^2 + m^2}$ with $%
m=-2G\langle\bar{q}q\rangle$ being the dynamically generated quark mass. The
density of states $\rho(k)$ is given by 
\begin{eqnarray}
\rho(k) = \frac{k^2}{2\pi^2}.
\end{eqnarray}

\begin{figure}\leavevmode
\begin{center}
\epsfxsize=8cm
\epsfbox{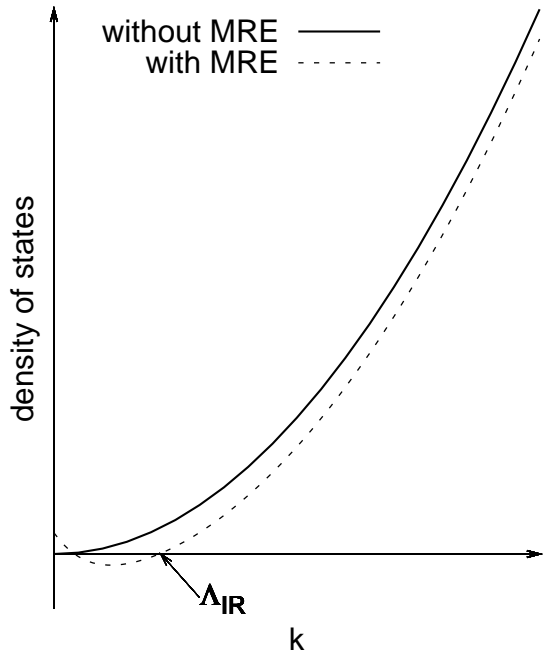}
\caption{The momentum dependence of the MRE density of states.}
\label{fig1}
\end{center}
\end{figure}

Now we consider a quark matter confined in a given sphere that is embedded
in the hadronic vacuum and do not impose the pressure balance condition
between interior of the sphere and external vacuum. We would rather fix the
radius of the sphere and examine the phase diagram. To incorporate
finite-size effects into the thermodynamic potential, we replace the density
of states $\rho (k)$ with the MRE density of states $\rho_{%
\mbox{\scriptsize
MRE}}(k,m,R)$, 
\begin{eqnarray}
\rho_{\mbox{\scriptsize MRE}}(k,m,R) = \frac{k^2}{2\pi^2} \left[1 + \frac{%
6\pi^2}{k R}f_S\left(\frac{k}{m}\right) + \frac{12 \pi^2}{(kR)^2} f_C\left(%
\frac{k}{m}\right) \right],  \label{eqn:MRE}
\end{eqnarray}
where $R$ is the radius of the sphere and the functions, 
\begin{eqnarray}
f_S (x) &=& -\frac{1}{8\pi}\left( 1 - \frac{2}{\pi}\arctan x \right), \\
f_C (x) &=& \frac{1}{12\pi^2}\left[ 1-\frac{3}{2}x\left( \frac{\pi}{2} -
\arctan x \right) \right],
\end{eqnarray}
correspond respectively to the surface and the curvature contributions to
the fermionic density of states. The parameter $m$ in Eq. (\ref{eqn:MRE})
should be identified with the mass of quarks in the interior of the sphere.
It should be also noted that the precise expression of the function $f_C(x)$
has not been derived within the MRE framework. Here we used the assumption
proposed in Ref. \cite{Mad3}. We show the momentum dependence of the MRE
density of states in Fig. \ref{fig1}. One can see that the MRE density of
states is suppressed by the finite-size effects and the suppression is more
pronounced at low momenta. It can also be seen that the MRE density of
states takes unphysical negative values at low momenta because of the
neglect of the higher order terms in $1/R$. Thus, we shall introduce the
infrared cutoff $\Lambda_{\mbox{\scriptsize IR}}$ that is defined by the
largest zero of $\rho_{\mbox{\scriptsize MRE}}(\Lambda_{IR},m,R)$, as is
shown in Fig. \ref{fig1}.

\begin{figure}\leavevmode
\begin{center}
\epsfxsize=8cm
\epsfbox{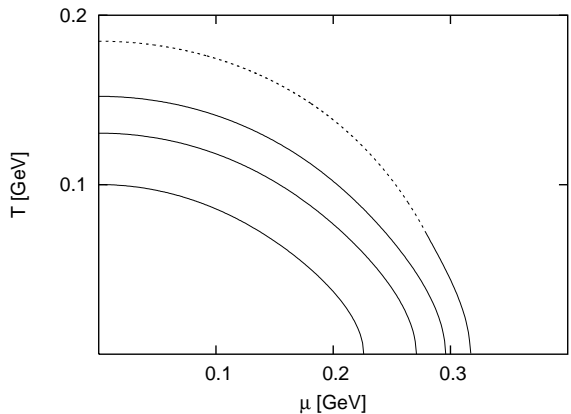}
\caption{The QCD phase diagram calculated using the MRE density of states
given in Eq. (3). The solid and dotted lines denote the first order and the
second order phase transition, respectively. Each line corresponds to $%
R=\infty$, $100$, $50$ and $30$ fm from the most outside one, respectively. }
\label{fig2}
\end{center}
\end{figure}

Thus, within the MRE approximation, the thermodynamic potential including
the finite-size effects is expressed as follows, 
\begin{eqnarray}
\omega = \frac{m^2}{4G}-\nu\int^{\Lambda_{\mbox{\scriptsize UV}}}_{\Lambda_{%
\mbox{\scriptsize IR}}} dk\rho_{\mbox{\scriptsize MRE}}\left\{E_k + \frac{1}{%
\beta}\ln\left[1+e^{-\beta(E_k + \mu)}\right] \left[1 + e^{-\beta(E_k - \mu)}%
\right] \right\}.  \label{eqn:TDP2}
\end{eqnarray}

To obtain the phase diagram we minimize Eq. (\ref{eqn:TDP2}) with respect to 
$m$. When the value of $m$ at the minimum is zero, the system is in the
chiral symmetric phase (QGP) and if the value of $m$ is not zero, the system
is in the broken symmetry phase (hadronic). We then find the phase diagram
given in Fig. \ref{fig2}. The outermost curve corresponds to $R=\infty $
(thermodynamical limit). In this limit, of course our approach recovers the
known results\cite{AY}. The order of the chiral phase transition is second
order at small chemical potential and as the chemical potential increases,
the order of the phase transition is changed form second to first at the
tricritical point. In this curve the parts corresponding to the second order
and the first order transitions are indicated by the dotted and solid lines,
respectively.

To see the effect of finite size in this approach, we show the critical
temperature lines for $R=$ $100$, $50$, and $30$ fm (from outside to
inside). One can see that the critical temperatures are lowered as $R$ is
decreased. This behavior is consistent with the enhancement of large
fluctuations induced by the finite-size effects. However, we note that the
tricritical point which exists at $R=\infty $ disappears for all finite
values of $R.$ At any value of chemical potential, the order of phase
transition becomes the first one and the second order phase transition
disappears. This is because, as we will see below, the MRE cannot describe
the second order phase transition.

These results obtained by the simple application of the MRE to the
thermodynamic potential are physically unreasonable for the following
reasons. First, the reduction of the critical temperature seems to be too
strong. Second, more critical one from physical point of view, is the
disappearance of the second order phase transition. In general, large
fluctuations of the order parameter could increase the order of the phase
transition, for instance, from first to second. Therefore, it is unnatural
that the second order phase transition was converted to the first order one.
This unphysical behavior indicates the necessity of some modification of the
present simple application of the MRE to the NJL model.

\begin{figure}\leavevmode
\begin{center}
\epsfxsize=8cm
\epsfbox{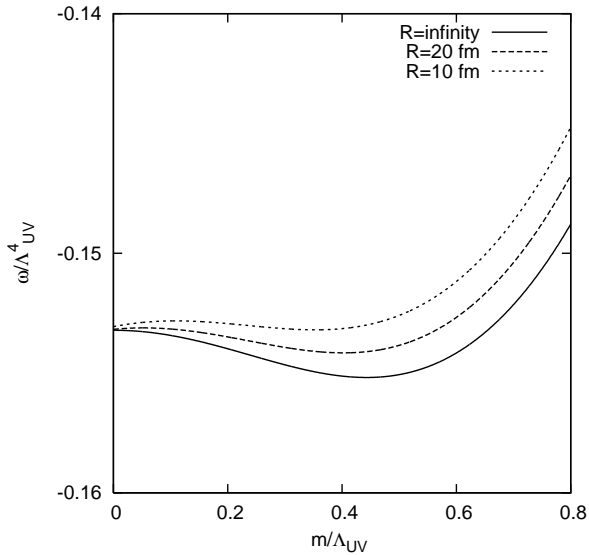}
\caption{The R dependence of the thermodynamic potential. The gradient at $%
m=0$ is positive when the finite-size effects are taken into account. }
\label{fig3}
\end{center}
\end{figure}

To clarify the problem mentioned above, let us take a closer look on the
self-consistency condition (SCC) including the MRE density of states. The
derivative of $\omega $ with respect to $m$ is given by 
\begin{eqnarray}
\frac{\partial \omega }{\partial m} &=&\frac{m}{2G}-\nu \int_{\Lambda _{%
\mbox{\scriptsize IR}}}^{\Lambda _{\mbox{\scriptsize UV}}}dk\frac{\partial
\rho _{\mbox{\scriptsize MRE}}}{\partial m}\left\{ E_{k}+\frac{1}{\beta }\ln %
\left[ 1+e^{-\beta (E_{k}+\mu )}\right] \left[ 1+e^{-\beta (E_{k}-\mu )}%
\right] \right\}  \nonumber \\
&&-\nu \int_{\Lambda _{\mbox{\scriptsize IR}}}^{\Lambda _{%
\mbox{\scriptsize
UV}}}dk\rho _{\mbox{\scriptsize MRE}}\frac{m}{E_{k}}\left[ 1-\frac{1}{%
e^{\beta (E_{k}+\mu )}+1}-\frac{1}{e^{\beta (E_{k}-\mu )}+1}\right]
\label{eqn:mscc}
\end{eqnarray}%
[the term proportional to 
$\partial \Lambda _{\mbox{\scriptsize IR}}/\partial m$ 
vanishes by virtue of $\rhom(\Lambda _{\mbox{\scriptsize IR}},m,R)=0$]. 
For the second order phase transition to exist, 
it is necessary that this derivative
vanishes at $m=0.$ However, as has been pointed out in Ref. \cite{KH}, $m=0$
is not the solution to the SCC because of 
\begin{equation}
\lim_{m\rightarrow 0}\frac{\partial \rho _{\mbox{\scriptsize MRE}}}{\partial
m}<0,
\end{equation}
so that 
\begin{equation}
\left. \frac{\partial \omega }{\partial m}\right\vert _{m=0}>0
\end{equation}
except for $R=\infty .$ That is, the thermodynamic potential has always
positive gradient at $m=0$, hence the thermodynamic potential (\ref{eqn:TDP2}%
) cannot describe the second order phase transition. In Fig. \ref{fig3}, the
thermodynamic potential at $(T,\mu )=(100,0)$ MeV is plotted as a function
of $m$ for $R=\infty $, $20$ and $10$ fm.

\section{Modification of the density of states}

In the proceeding section, we examined the $T$-$\mu$ phase diagram, looking
for the minimum of the thermodynamic potential. This sort of the application
of the MRE density of states has been employed in Refs. \cite{KH,YHT}. The
second order phase transition suddenly disappears soon after taking the
finite-size effect into account. Thus, the phase diagram with finite
size does not converge to that with infinite size even if the radius $R$
is increased. It follows that we cannot take the thermodynamic limit in the
MRE approximation. Moreover, fluctuations due to the finite-size effects is
expected to change the order of phase transition from first to second.
Hence, it is unacceptable that the finite size effect is taken account of
the MRE approximation employed in the proceeding section. In this section,
we reconsider the application of the MRE to resolve this discrepancy.

The reason why the second order chiral phase transition disappeared in the
previous section is the nonexistence of the trivial solution to the SCC. The
disappearance is due to the mass dependence contained in the MRE density of
states which prevents the disappearance of the second order phase transition
is due to the mass dependence contained in the MRE density of states. As is
shown in the last of the proceeding section, the SCC does not have the
trivial solution because of the mass dependence.

However, there can be another scenario for the application of the MRE.
According to the paper by Bailin and Bloch\cite{BB}, the parameter contained
in the MRE density of states is not the mass of particle inside the cavities
but a parameter which characterizes the penetration of the wave function of
the quark matter confined in a finite volume into the hadronic environment.
In the following, we impose specific boundary conditions on quarks and study
the cases in which the MRE density of states does not depend on the
dynamical quark mass. In this scenario, the mass parameter contained in the
MRE density of states is replaced by the parameter $\kappa$ which is
generally independent of the dynamical quark mass. Then, the SCC is given by 
\begin{eqnarray}
\frac{m}{2G} = \nu\int^{\Lambda_{\mbox{\scriptsize UV}}}_{\Lambda_{%
\mbox{\scriptsize IR}}}dk \rho_{\mbox{\scriptsize MRE}}(k,\kappa,R) \frac{m}{%
E_k} \left[1-\frac{1}{e^{\beta(E_k + \mu)}+1} +\frac{1}{e^{\beta(E_k -
\mu)}+1} \right].
\end{eqnarray}
We do not have the derivative term of the MRE density of states and the SCC
obviously has the trivial solution.

\begin{figure}\leavevmode
\begin{center}
\epsfxsize=8cm
\epsfbox{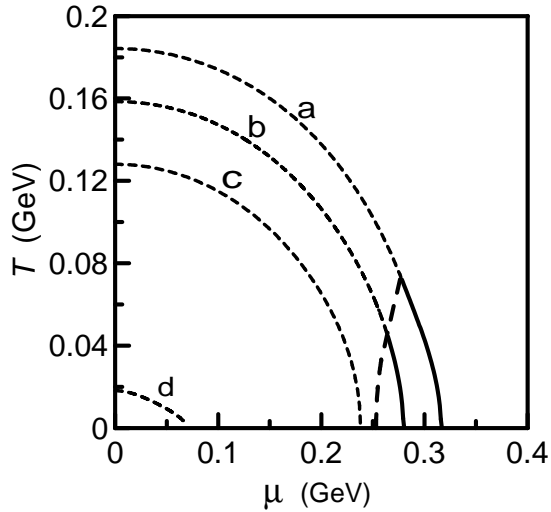}
\caption{The QCD phase diagram calculated using the MRE density of states
given in Eq. (11) with $\protect\kappa=\infty$. The solid and dotted lines
represents the first order and second order phase transition, respectively.
Each line corresponds to (a) $R=\infty$, (b) $20$, (c) $10$ and (d) $6$ fm 
from the most outside
one, respectively. The dashed represents the trajectory of the
tricritical point as a function of $R$. 
At the radius smaller than $R=5.7$ fm, 
our calculation does not show the chiral symmetry breaking.}
\label{fig4}
\end{center}
\end{figure}

For instance, let us consider the Dirichlet boundary condition by setting $%
\kappa =\infty $\cite{BB}. Then, we obtain 
\begin{eqnarray}
\lim_{\kappa \rightarrow \infty }f_{S}(k/\kappa ) &=&-\frac{1}{8\pi }, \\
\lim_{\kappa \rightarrow \infty }f_{C}(k/\kappa ) &=&\frac{1}{12\pi ^{2}},
\end{eqnarray}%
and the infrared cutoff is given by 
\begin{equation}
\Lambda _{\mbox{\scriptsize IR}}=\frac{3\pi +\sqrt{9\pi ^{2}-64}}{8R}\simeq 
\frac{1.8}{R}.
\end{equation}

In Fig. \ref{fig4}, we show the corresponding phase diagram. The solid and
dotted lines represents the critical lines of the first order and second
order phase transitions, respectively. These lines represent the critical
lines of (a) $R=\infty $, (b) $20$, (c) $10$ and (d) $5.7$ fm from the out-most one, respectively.
Compared to the proceeding result, the finite-size effects are suppressed.

The dashed line represents the trajectory of the tricritical point as a
function of $R$. One can easily see that the temperature and the chemical
potential of the tricritical point are reduced as the $R$ is decreased. 
Finally, the tricritical point disappears at $R=12$ fm and hence
the region of the first order phase transition vanishes. It
should be noted that the phase diagram calculated with the MRE density of
states converges to that in the infinite size in the limit of $R
\longrightarrow \infty$, and is consistent with the fact of the increase of
fluctuations due to the finite size effects. Thus, this approach seems to be
more plausible than the proceeding approximation.

Note that even the second order phase transition disappears at the radius smaller than $R=5.7$ fm.
This is because of the existence of the infrared cutoff and we will find the phase transition at smaller $R$ 
by taking account of the higher order terms of the MRE density of state.

If this boundary condition is applicable for the real physical situation
created in relativistic heavy ion collisions, then the equation of state of
the matter changes with the expansion of the system. At the early stage of
creation of the hot matter at the center, the system size is the order of $%
10 $ fm and the critical temperature at $\mu =0$ is around $130$ MeV, but
after the system expands and reaches the hadronic freezeout stage, then the
system size is much larger and the critical temperature almost recovers the
thermodynamical limit. Therefore, the hydrodynamical description of such a
situation requires an equation of state which reflects the above finite-size effect.

The physical environment consistent with the MIT bag model is considered to
be realized by setting the Neumann boundary condition, $\kappa=0$, 
\begin{eqnarray}
\lim_{\kappa \rightarrow 0}f_S (k/\kappa) &=& 0, \\
\lim_{\kappa \rightarrow 0}f_C (k/\kappa) &=& -\frac{1}{24\pi^2},
\end{eqnarray}
and the infrared cut off is 
\begin{eqnarray}
\Lambda_{\mbox{\scriptsize IR}} = \frac{1}{\sqrt{2}R} .
\end{eqnarray}

\begin{figure}\leavevmode
\begin{center}
\epsfxsize=8cm
\epsfbox{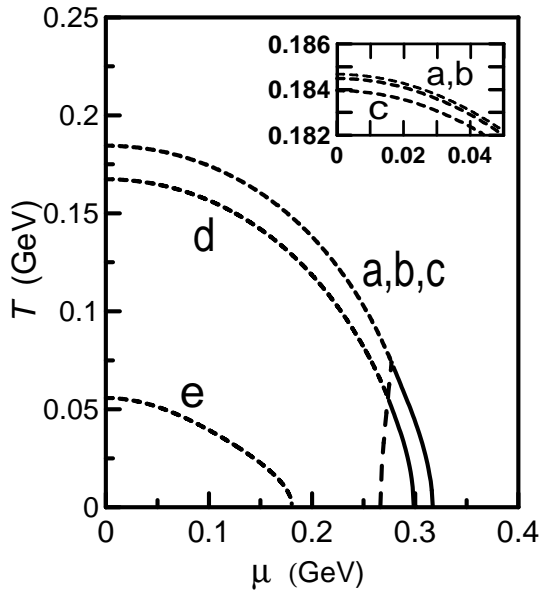}
\caption{The QCD phase diagram calculated using the MRE density of states
given in Eq. (11) with $\protect\kappa = 0$. The solid and dotted lines
represents the first order and second order phase transition, respectively.
Each line corresponds to (a) $\infty$, (b) $20$, (c) $10$, (d) $2$ and (e) $0.9$ fm 
from the most
outside one, respectively. The three lines corresponding to $R=\infty$, $20$
and $10$ fm almost degenerate. The dashed represents the trajectory of
the tricritical point as a function of $R$. 
At the radius smaller than $R=0.9$ fm, 
our calculation does not show the chiral symmetry breaking.
}
\label{fig5}
\end{center}
\end{figure}

In Fig. \ref{fig5}, we show the corresponding phase diagram. The lines
correspond to the critical lines for the cases of (a) $R=\infty$, (b) $20$, (c) $10$, (d) $2$ 
and (e) $0.9$ fm from the out-most one, respectively. The finite-size effects are
more suppressed, because the surface term vanishes in this boundary
condition, and hence the difference of the critical lines between $R=\infty$
and $R=20$ fm in Fig. \ref{fig4} is negligible. The remarkable reduction of
the critical temperature is observed at smaller size than $R=2$ fm. The
trajectory of the tricritical point denoted by the dashed line changes
steeper than the case of the Dirichlet boundary condition.
In this boundary condition, the tricritical point vanishes at $R=1.2$ fm and 
the second order phase transition disappears at the radius smaller than $R=0.9$ fm.

Contrary to the previous case, results from this boundary condition indicate
that the finite size system with $R=20$ fm is large enough to take the
thermodynamic limit. In this case, it will be reasonable to apply the usual
equation of state which is given by the thermodynamical limit in
hydrodynamical models to describe the heavy-ion collision processes.

So far, we have discussed the behavior of the thermodynamic potential.
However, it is not easy to see how the second order phase transition is
affected by the finite-size effects. For the purpose of looking at
characteristic changes of the second order transition, we calculate the
specific heat defined by 
\begin{eqnarray}
C &=& -T\left(\frac{\partial^2 \Omega}{\partial T^2}\right)_{\mu}.
\end{eqnarray}
Note that the specific heat shows discontinuity at the critical point of the
second order phase transition.

Figures 6 and 7 show the specific heat of the systems for the cases of $%
\kappa = 0$ and $\kappa = \infty$, respectively. Here, we chose $\mu=0.1$
GeV for a common quark chemical potential and examined the cases of $R=100$
fm and $R=10$ fm. In both cases, the specific heats jump at the
corresponding critical temperatures. The jump of the specific heat still
exists after including the finite-size effects. The result indicates that
the second order phase transitions remain second order. However, the gap of
the specific heat is decreased by the finite-size effects.

\begin{figure}\leavevmode
\begin{center}
\epsfxsize=8cm
\epsfbox{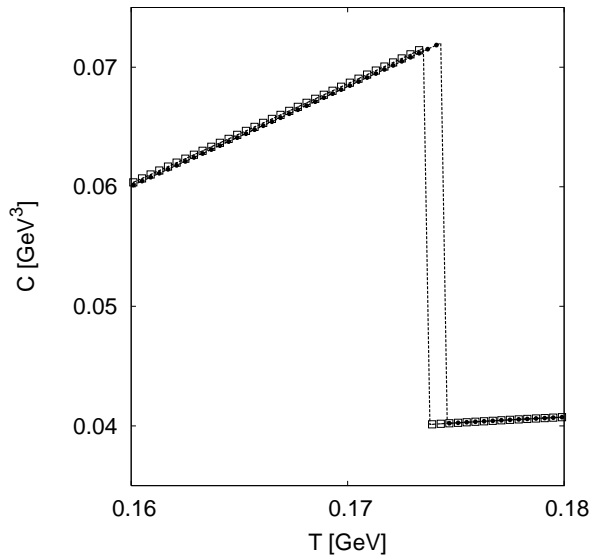}
\caption{ The specific heat $C$ as a function of $T$ for the cases of $R=100$
fm ($\bullet$) and $R=10$ fm ($\Box$). The mass parameter is taken to be $%
\protect\kappa=0$. }
\label{fig7}
\end{center}
\end{figure}

\begin{figure}\leavevmode
\begin{center}
\epsfxsize=8cm
\epsfbox{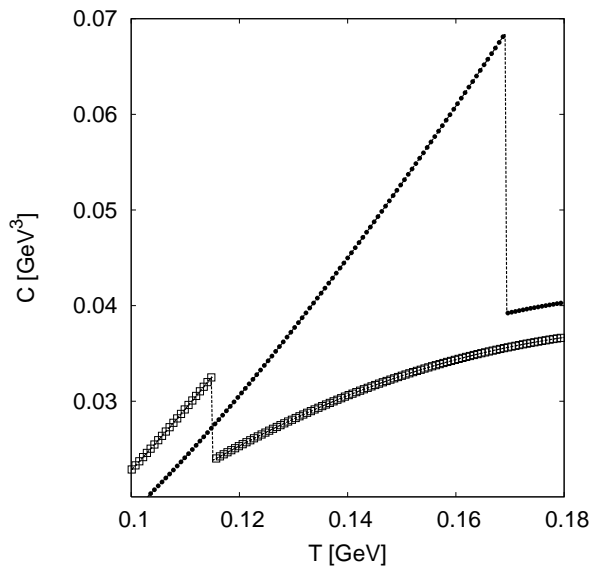}
\caption{The specific heat $C$ as a function of $T$ for the cases of $R=100$
fm ($\bullet$) and $R=10$ fm ($\Box$). The mass parameter is taken to be $%
\protect\kappa=\infty$. }
\label{fig8}
\end{center}
\end{figure}

\section{summary and concluding remarks}

In this paper, wee have discussed the finite-size effects on the chiral
phase transition in QCD in the context of the two-flavor Nambu-Jona-Lasinio
model in the mean-field approximation. To incorporate finite-size effects
into the thermodynamic potential, we employed the multiple reflection
expansion (MRE), where finite-size effects are included by the modification
of the density of state.

We applied the MRE approximation in different two ways. One is that the
\textquotedblleft mass\textquotedblright\ parameter contained in the MRE
density of state is identified as the dynamical quark mass as is done in 
\cite{KH,YHT}. In this case it is shown that only the first order phase
transition can take place for any finite size and the critical point
disappears.

The other is that we identify the \textquotedblleft mass\textquotedblright\
parameter as the inverse of the logarithmic derivative of the wave function
at the surface,  following the discussion by Balian and Bloch\cite{BB}. This
corresponds to the situation that our system is confined in a finite volume
by an external potential and the boundary condition is fixed independently
from the order parameter of the system.. Then, the order of chiral phase
transition changes from the second to the first as the chemical potential
increases. The temperature and the chemical potential of at the critical
point become smaller as the system size decreases. This behavior is
consistent with the fact that fluctuations is enhanced by the finite-size
effects. However, quantitatively, we found that the two extreme boundary
conditions lead to substantially different results. For the Dirichlet
boundary condition, the finite-size effect is still large. If this is the
case, the critical temperature changes from $T_{c} \sim 130$ MeV to 
$T_{c} \sim 180$ MeV at zero chemical potential if the size of the
system expands from $10$ fm to $50$ fm. Such effects should reflect on the
hydrodynamical collective observables, i.e., the $v_{2}$. On the other hand,
for the Neumann boundary condition, the finite-size effects are not so
visible and the equation of state is almost the same as that obtained from
the thermodynamical limit.

It should be noted that the MRE calculation will lose its validity when the
size of our system is very small since we have ignored the higher order
terms of the MRE approximation. Moreover, the finite-size effects
considered in the MRE approximation is only the reduction of the density of
state. However, in general, we can expect other influences due to the finite-size 
effects; for example, the energy eigenvalue of particles will be also
affected. 

We calculated the thermodynamic potential in the mean-field approximation,
although it is known that large fluctuations near phase transitions
invalidate the mean-field approximation. However, the QCD critical
temperature in the mean-field approximation is qualitatively coincide with
that of the lattice QCD calculations. Thus, the results obtained in this
work would be still qualitatively reliable, unless the fluctuations are
extraordinary enhanced by the finite-size effects.

Finally, we comment on the outlook for future studies. In this paper, we
studied the static behavior of chiral symmetry. However, the finite-size
effects would also affect the dynamical behaviors. For instance, the life
time of fluctuations near phase transitions increases because of the
critical slowing down (CSD)\cite{Koide1,Koide2}. The CSD is due to the large
fluctuations, and hence the finite-size effects will enhance the CSD.

\begin{acknowledgments}
The authors thank C. E. Aguiar and T. H. Elze for fruitful discussion and comments. 
The authors acknowledge 
to financial support of FAPESP, FAPERJ, CNPq and CAPES/PROBRAL.
\end{acknowledgments}

\end{document}